\documentclass[aps]{revtex4}
\usepackage{amsmath}
\usepackage{graphicx}
\usepackage{slashed}
\begin{document}
\title{Critical Scaling and a Dynamical Higgs Boson}
\author{Philip D. Mannheim\\
Department of Physics, University of Connecticut, Storrs, CT 06269, USA\\
email: philip.mannheim@uconn.edu}
\date{February 5, 2017}

\begin{abstract}
In  a quantum electrodynamics theory that is realized by critical scaling and anomalous dimensions, the action is not chiral invariant and there are no dynamical Goldstone or Higgs boson bound states. In the mean-field approximation to a chiral invariant four-fermion theory the associated mean-field sector action is not chiral invariant either and it also possesses no dynamical bound states, with Goldstone and Higgs bosons instead being generated by an accompanying four-fermion residual interaction. In this paper we show that if a critical scaling electrodynamics in which the dimension $d_{\theta}$ of $\bar{\psi}\psi$ is reduced from three to two is augmented with a four-fermion interaction, precisely because it possesses no dynamical bound states the electrodynamic sector can be  reinterpreted as a mean-field approximation to a larger theory that is chiral symmetric. And with $d_{\theta}=2$ we show in this larger theory there is a residual interaction that then does generate dynamical Goldstone and Higgs bosons in scattering amplitudes that are completely finite. While the dynamically generated Goldstone boson is massless, the dynamically induced Higgs boson is found to be a narrow resonance just above threshold, with its width being a diagnostic that could potentially enable one to distinguish between a dynamical Higgs boson and an elementary one. 
\end{abstract}
\maketitle

\section{Introduction}
With the discovery \cite{ATLAS2012,CMS2012} of the Higgs boson associated with the Higgs mechanism \cite{Englert1964,Higgs1964a,Higgs1964b,Guralnik1964}, it has become imperative to determine whether the Higgs boson is an elementary field that appears in the fundamental Lagrangian of nature, or whether it is a fermion-antifermion composite bound state. If the Higgs boson is to be elementary, one has to explain why its quadratically divergent self-energy is not regularized at a very high grandunified or Planck mass scale (the hierarchy problem), and one has to explain why the cosmological constant term that the Higgs field induces when it acquires a non-vanishing vacuum expectation value is not at the same mass scale as the 125 GeV mass that the Higgs boson is now known to have. If the Higgs boson is to be composite, one needs to develop a suitable dynamical mass generation scheme in a renormalizable quantum field theory that would not only not lead to an unacceptably high mass for the Higgs boson or an unacceptably large cosmological constant term, but would recover the successful standard phenomenology associated with an elementary Higgs boson. Here we present a dynamical scheme that does this, a scheme that does not differ from an elementary Higgs boson theory when the Higgs boson is off shell (off-shell Higgs exchange and loop diagrams), but does differ from it on the mass shell. On shell the dynamical scheme requires that the Higgs boson be a narrow resonance just above threshold, with its width being a diagnostic that could potentially distinguish a dynamical Higgs boson from an elementary one. 

The dynamical scheme that we study is a quantum electrodynamics (QED) theory as realized by critical scaling with anomalous dimensions in the ultraviolet, and by the 
non-vanishing of the vacuum expectation value of the composite operator $\bar{\psi}\psi$ due infrared dynamics, with the dynamical dimension of $\bar{\psi}\psi$ being reduced from three to two. It has long been known that because of the Baker-Johnson mass renormalization anomaly \cite{Baker1971a}, if such critical scaling is to occur no dynamical pseudoscalar Goldstone or scalar Higgs boson is generated, with the chiral symmetry being broken in the Lagrangian, and with the theory actually corresponding to a theory with an intrinsic mass with Lagrangian ${\cal{L}}_{\rm QED}^m={\cal{L}}_{\rm QED}^0-m\bar{\psi}\psi$ where ${\cal{L}}^0_{\rm QED}=-(1/4)F_{\mu\nu}F^{\mu\nu}+\bar{\psi}\gamma^{\mu}(i\partial_{\mu}-eA_{\mu})\psi$. The fact that there would be no composite bound states in a critical scaling QED has a parallel in the chiral-invariant Nambu-Jona-Lasinio (NJL) four-fermion model \cite{Nambu1961},  where the Lagrangian is broken into two pieces, a mean-field piece and a residual interaction, with the mean-field sector piece containing a mass term that is not present in the original NJL Lagrangian. Neither of these two pieces is separately chirally symmetric, with the composite bound states being found not in the mean-field sector at all but in the residual interaction sector. Thus by augmenting a  massless fermion ${\cal{L}}^0_{\rm QED}$ with a four-fermion interaction, we can recognize an ${\cal{L}}^m_{\rm QED}$ with its fermion mass term as a mean-field theory, and as such this sector should not have any composite bound states since a mean-field sector never does. However, dynamical Goldstone and Higgs boson bound states are generated by the four-fermion residual interaction. And with $(\bar{\psi}\psi)^2$ acting as an operator whose dynamical dimension has been reduced from six to four, the four-fermion sector is power-counting renormalizable, and the mass of the ensuing Higgs boson is found to be both finite and at the same mass scale as that of the dynamically generated fermion rather than at a high regulator mass scale. By coupling the theory to an equally conformal invariant gravity theory (conformal gravity), the interplay between the gravity and four-fermion sector is found to take care of the cosmological constant that is induced by the dynamical symmetry breaking procedure.

\section{The NJL Chiral Four-Fermion Model}

\subsection{NJL Model as a Mean-Field Theory}

\begin{figure}[htpb]
\begin{center}
\includegraphics[width=3.3in,height=1.3in]{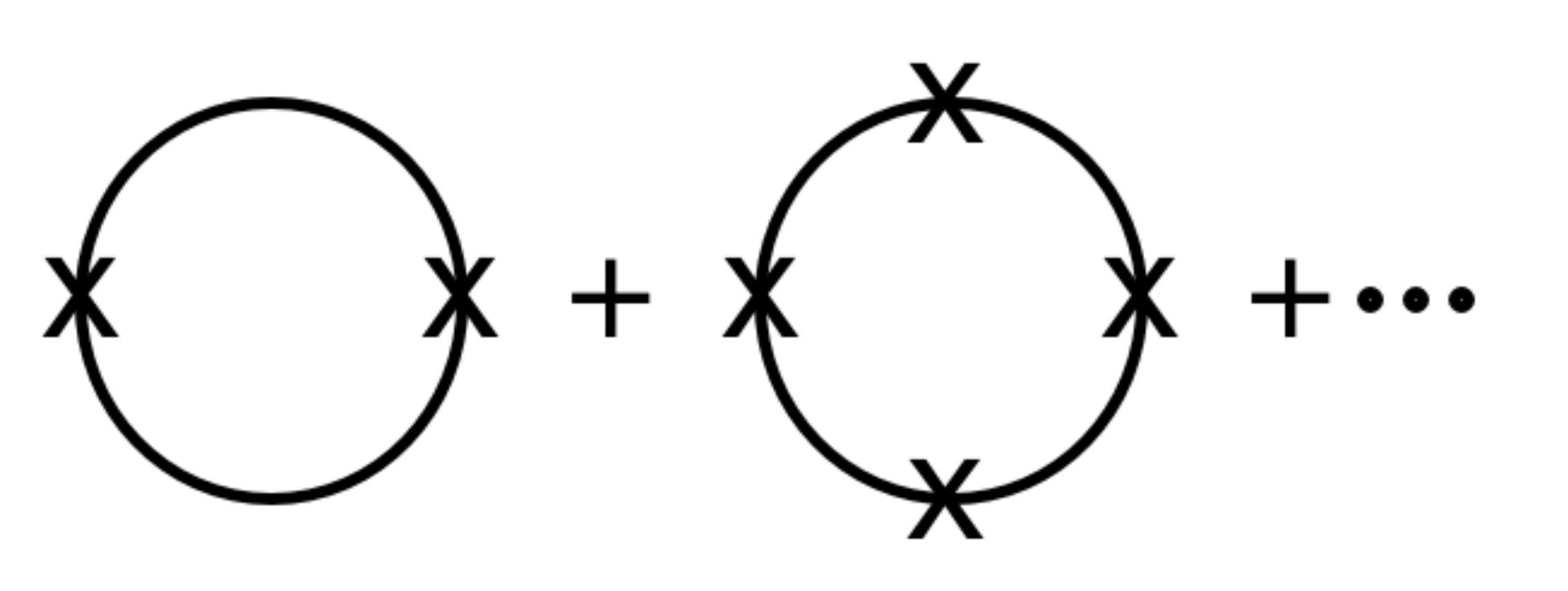}
\end{center}
\caption{Vacuum energy density $\epsilon(m)$ via an infinite summation of massless graphs with zero-momentum point $m\bar{\psi}\psi$  insertions.}
\label{fig1}
\end{figure}

Since our interest in this paper is in studying critical scaling in a four-fermion theory, we will need to generalize the standard analysis of the NJL model. In fact our approach will be seen to lead us to a renormalizable version of the NJL model. The NJL  action
\begin{eqnarray}
I_{\rm NJL}=\int d^4x \left[i\bar{\psi}\gamma^{\mu}\partial_{\mu}\psi-\frac{g}{2}[\bar{\psi}\psi]^2-\frac{g}{2}[\bar{\psi}i\gamma^5\psi]^2\right]
\label{L1a}
\end{eqnarray}
is invariant under chiral  transformations of the form $\psi(x)\rightarrow \exp(i\alpha\gamma^5)\psi(x)$, with the chiral symmetry preventing the presence of any fermion mass term in the action. Nonetheless, a fermion mass can be induced dynamically in the model. A direct way to show this is to introduce a mass term, and break the action into mean-field and residual interaction pieces according to $I_{\rm NJL}=I_{\rm MF}+I_{\rm RI}$, where
\begin{eqnarray}
I_{\rm MF}=\int d^4x \left[i \bar{\psi}\gamma^{\mu}\partial_{\mu}\psi-m\bar{\psi}\psi +\frac{m^2}{2g}\right],\qquad
I_{\rm RI}=\int d^4x \left[-\frac{g}{2}\left(\bar{\psi}\psi-\frac{m}{g}\right)^2-\frac{g}{2}\left(\bar{\psi}i\gamma^5\psi\right)^2\right].
\label{L2a}
\end{eqnarray}
In this decomposition neither $I_{\rm MF}$ nor $I_{\rm RI}$ is separately chiral invariant, only their sum is. One then treats $m$ as a trial parameter and looks to show that a state $|\Omega_m\rangle$ with non-zero $m$ has lower energy than the state $|\Omega_0\rangle$  with $m=0$,  to thus be energetically favored. In the Hartree-Fock approximation one sets
\begin{eqnarray}
\langle \Omega_{\rm m}|\left[\bar{\psi}\psi-\frac{m}{g}\right]^2|\Omega_{\rm m}\rangle=\langle \Omega_{\rm m}|\left[\bar{\psi}\psi-\frac{m}{g}\right]|\Omega_{\rm m}\rangle^2=0,\qquad \langle \Omega_{\rm m}|\bar{\psi}i\gamma^5\psi|\Omega_{\rm m}\rangle=0,
\label{L3a}
\end{eqnarray}
\begin{eqnarray}
\langle \Omega_{\rm m}|\bar{\psi}\psi|\Omega_{\rm m}\rangle=-i\int \frac{d^4k}{(2\pi)^4} {\rm Tr}\left[\frac{1}{\slashed{p}-m+i\epsilon}\right]=
\frac{m}{g}, 
\label{L4a}
\end{eqnarray}
and defines the physical mass $m$ to be that value of $m$ that satisfies $\langle \Omega_{\rm M}|\bar{\psi}\psi|\Omega_{\rm M}\rangle=M/g$, i.e. which satisfies the gap equation
\begin{eqnarray}
-\frac{M\Lambda^2}{4\pi^2}+\frac{M^3}{4\pi^2}{\rm ln}\left(\frac{\Lambda^2}{M^2}\right)=\frac{M}{g},
\label{L5a}
\end{eqnarray}
where $\Lambda$ is a cutoff that is required since the NJL model is not renormalizable. The energy-density difference in a volume $V$, viz. $\epsilon(m)=(\langle \Omega_m|H|\Omega_m\rangle-\langle \Omega_0|H|\Omega_0\rangle)/V$, 
associated with $I_{\rm MF}$ is given (see e.g. \cite{Mannheim2016}) by the infinite sum $\sum(1/n!)G^{(n)}_0(q_{\mu}=0,m=0)m^n$  of massless  fermion graphs with point (i.e. not dressed) $m\bar{\psi}\psi$ insertions as exhibited in Fig. (\ref{fig1}), to yield 
an $\tilde{\epsilon}(m)=\epsilon(m)-m^2/2g$ of the form
\begin{eqnarray}
\tilde{\epsilon}(m)=i\int \frac{d^4p}{(2\pi)^4}{\rm Tr}\bigg{[}{\rm ln}\left(\frac{\slashed{p}-m+i\epsilon}{\slashed{p}+i\epsilon}\right)\bigg{]}-\frac{m^2}{2g}~~
=\frac{m^4}{16\pi^2}{\rm ln}\left(\frac{\Lambda^2}{m^2}\right)-\frac{m^2M^2}{8\pi^2}{\rm ln}\left(\frac{\Lambda^2}{M^2}\right)+\frac{m^4}{32\pi^2},~~
\label{L6a}
\end{eqnarray}
with $\tilde{\epsilon}(M)$ indeed being negative, just as required.

\subsection{Higgs-Like Lagrangian}

\begin{figure}[htpb]
\begin{center}
\includegraphics[width=3.4in,height=0.8in]{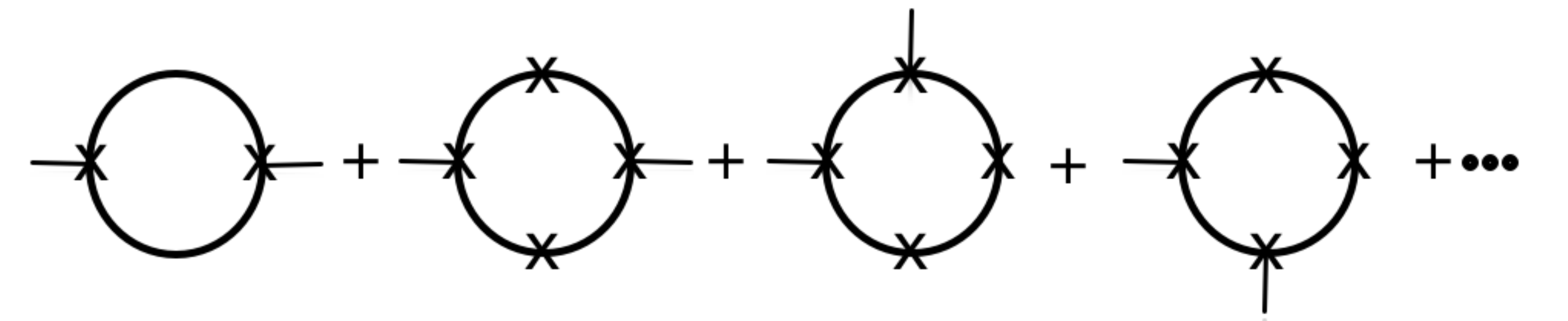}
\end{center}
\caption{$\Pi_{\rm S}(q^2,m(x))$ developed as an infinite summation of massless graphs, each with two point $m\bar{\psi}\psi$ insertions carrying momentum $q_{\mu}$ (shown as external lines), with all other point $m\bar{\psi}\psi$ insertions carrying zero momentum.}
\label{fig2}
\end{figure}

To generate a spacetime dependence to $m(x)$, we introduce a spacetime-dependent coherent state $|C\rangle$ and consider spacetime-dependent matrix elements of the form $\langle C|\bar{\psi}(x)\psi(x)|C\rangle$. For $\tilde{\epsilon}(m)$ we can replace $m$ by $m(x)$, since even though $m(x)$ is not spacetime independent, the set of graphs given in Fig. (\ref{fig1}) is a set of graphs in each one of which the fermion is massless, so that $m(x)$ does not appear inside the individual momentum integrals, but only multiplies them by $m^n(x)$. (We thus do the momentum integrals with constant $m$ first and then set $m=m(x)$ afterwards.) For a kinetic energy term for $m(x)$ we introduce $\Pi_{\rm S}(q^2, m)$ as defined in (\ref{L9a}) below, and evaluate $\Pi_{\rm S}(q^2, m(x))$ as \cite{Mannheim1976} the series of massless graphs given in Fig. (\ref{fig2}), where again $m(x)$ does not appear inside the momentum integrals. We can then identify $-d\Pi_{\rm S}(q^2, m(x))/dq^2|_{q^2=0}=[{\rm ln}(\Lambda^2/m^2(x))-5/3]/8\pi^2$ as the coefficient of $(1/2)\partial_{\mu}m(x)\partial^{\mu}m(x)$, to obtain a leading cutoff dependent effective action of the form \cite{Eguchi1974,Mannheim1976}
\begin{eqnarray}
I_{\rm EFF}=\int \frac{d^4x}{8\pi^2}{\rm ln}\left(\frac{\Lambda^2}{M^2}\right)\bigg[
\frac{1}{2}\partial_{\mu}m(x)\partial^{\mu}m(x)
+m^2(x)M^2-\frac{1}{2}m^4(x)\bigg].
\label{L7a}
\end{eqnarray}
If we make the chiral symmetry local and introduce  a coupling $g_{\rm A}\bar{\psi}\gamma^{\mu}\gamma^5A_{\mu 5}\psi $ to an axial gauge field $A_{\mu 5}(x)$, on setting $\phi(x)/g=\langle C|\bar{\psi}(x)(1+\gamma^5)\psi(x)|C\rangle$ the effective action becomes the local \cite{Eguchi1974,Mannheim1976}
\begin{eqnarray}
I_{\rm LOC}=\int \frac{d^4x}{8\pi^2}{\rm ln}\left(\frac{\Lambda^2}{M^2}\right)\bigg[
\frac{1}{2}|(\partial_{\mu}-2ig_{\rm A}A_{\mu 5})\phi(x)|^2
+|\phi(x)|^2M^2-\frac{1}{2}|\phi(x)|^4-\frac{g_{\rm A}^2}{6}F_{\mu\nu  5}F^{\mu\nu}_{5}\bigg].
\label{L8a}
\end{eqnarray}
We recognize this action as a double-well Ginzburg-Landau type Higgs Lagrangian with order parameter $\phi(x)$, only now generated dynamically.

In $I_{\rm EFF}$ of (\ref{L7a}) there is a double-well Higgs potential, but since $m(x)/g=\langle C|\bar{\psi}(x)\psi(x)|C\rangle$ is a c-number, $m(x)$ does not itself represent a q-number scalar field. Rather, as we now show, the q-number fields are to be found as collective bound state modes generated by the residual interaction, with no fundamental scalar fields being needed at all. 

\subsection{The Collective Goldstone and Higgs Modes}

To identify dynamical bound states we introduce the scalar and pseudoscalar vacuum polarizations associated with $I_{\rm MF}$, viz. (cf. Fig. (\ref{fig2}) and its pseudoscalar analog) 
\begin{eqnarray}
\Pi_{\rm S}(q^2,M)&=&\int d^4xe^{iq\cdot x}\langle\Omega_M|\bar{\psi}(x)\psi(x)\bar{\psi}(0)\psi(0)|\Omega_M\rangle
=-i\int\frac{d^4p}{(2\pi)^4}{\rm Tr}\left[\frac{1}{\slashed{p}-M+i\epsilon}\frac{1}{\slashed{p}+\slashed{q}-M+i\epsilon}\right]
\nonumber\\
&=&
-\frac{\Lambda^2}{4\pi^2}
+\frac{M^2}{4\pi^2}{\rm ln}\left(\frac{\Lambda^2}{M^2}\right)
+\frac{(4M^2-q^2)}{8\pi^2} 
+\frac{(4M^2-q^2)}{8\pi^2}{\rm ln}\left(\frac{\Lambda^2}{M^2}\right)
\nonumber\\
&&-\frac{1}{8\pi^2}\frac{(4M^2-q^2)^{3/2}}{(-q^2)^{1/2}}
{\rm ln}\left(\frac{(4M^2-q^2)^{1/2}+(-q^2)^{1/2}}{(4M^2-q^2)^{1/2}-(-q^2)^{1/2}}\right),
\label{L9a}
\end{eqnarray}
\begin{eqnarray}
\Pi_{\rm P}(q^2,M)
&=&\int d^4xe^{iq\cdot x}\langle\Omega_M|\bar{\psi}(x)i\gamma^5\psi(x)\bar{\psi}(0)i\gamma^5\psi(0)|\Omega_M\rangle
=-i\int\frac{d^4p}{(2\pi)^4}{\rm Tr}\left[i\gamma^5\frac{1}{\slashed{p}-M+i\epsilon}i\gamma^5\frac{1}{\slashed{p}+\slashed{q}-M+i\epsilon}\right]
\nonumber\\
&=&-\frac{\Lambda^2}{4\pi^2}
+\frac{M^2}{4\pi^2}{\rm ln}\left(\frac{\Lambda^2}{M^2}\right) 
-\frac{q^2}{8\pi^2}{\rm ln}\left(\frac{\Lambda^2}{M^2}\right) 
+\frac{(4M^2-q^2)}{8\pi^2}
\nonumber\\
&&+\frac{(8M^4-8M^2q^2+q^4)}{8\pi^2 (-q^2)^{1/2}(4M^2-q^2)^{1/2}}
{\rm ln}\left(\frac{(4M^2-q^2)^{1/2}+(-q^2)^{1/2}}{(4M^2-q^2)^{1/2}-(-q^2)^{1/2}}\right).
\label{L10a}
\end{eqnarray}
The scattering matrices in the two channels are given by using $ I_{\rm RI}$ to iterate the vacuum polarizations according to $T=g+g\Pi g+g\Pi g\Pi g+...$, to yield
\begin{eqnarray}
T_{\rm S}(q^2,M)=\frac{1}{g^{-1}-\Pi_{\rm S}(q^2,M)},\qquad
T_{\rm P}(q^2,M)=\frac{1}{g^{-1}-\Pi_{\rm P}(q^2,M)}.
\label{L11a}
\end{eqnarray}
With $g^{-1}$ being given by $\langle \Omega_{\rm M}|\bar{\psi}\psi|\Omega_{\rm M}\rangle=M/g$, we find poles in each of the two channels, and near them the scattering matrices behave as 
\begin{eqnarray}
T_{\rm S}(q^2,M)&=&\frac{R_{\rm S}^{-1}}{(q^2-4M^2)},~~T_{\rm P}(q^2,M)=\frac{R_{\rm P}^{-1}}{q^2},
\label{L12a}
\end{eqnarray}
where $R_{\rm S}=R_{\rm P}={\rm ln}(\Lambda^2/M^2)/8\pi^2$.
Dynamical pseudoscalar Goldstone and scalar Higgs boson states are thus induced by the residual interaction even though neither is contained in the mean-field sector, with the mass of the scalar Higgs boson naturally being of order the mass of the fermion and not of order the cutoff $\Lambda$. The NJL model thus contains the key ingredients needed for dynamical symmetry breaking. However the model is not renormalizable. To make it renormalizable we turn now to critical scaling and anomalous dimensions, as this will serve to sufficiently soften the point four-fermion couplings.

\section{Critical Scaling in Quantum Electrodynamics}

\subsection{Vanishing of the Bare Fermion Mass}

In a study of quantum electrodynamics, Johnson, Baker, and Willey \cite{Johnson1964,Johnson1967,Baker1971a} found that the standard $Z_3$ and $\delta m$ renormalization constants of the theory would be finite in the event of critical scaling and anomalous dimensions. Their results can be summarized by noting that in the generalized Landau gauge the asymptotic renormalized fermion propagator $\tilde{S}^{-1}(p)$ obeys \cite{Adler1971} the Callan-Symanzik equation
\begin{eqnarray}
\left[m\frac{\partial}{m}+\beta(\alpha)\frac{\partial}{\partial \alpha}\right]\tilde{S}^{-1}(p)=m[\gamma_{\theta}(\alpha)-1]\tilde{\Gamma}_{\rm S}(p,p,0),
\label{L13a}
\end{eqnarray}
where $\tilde{\Gamma}_{\rm S}(p,p,0)$ is the renormalized Green's function associated with the insertion of a zero-momentum composite operator $\theta=\bar{\psi}\psi$ into the inverse fermion propagator. Critical scaling and the finiteness of $Z_3$ is achievable if $\beta(\alpha)=0$, with one then asymptotically having 
\begin{eqnarray}
\tilde{S}^{-1}(p)= \slashed{p}-m\left(\frac{-p^2-i\epsilon}{m^2}\right)^{\gamma_{\theta}(\alpha)/2},
\qquad
\tilde{\Gamma}_{\rm S}(p,p,0)=\left(\frac{-p^2-i\epsilon}{m^2}\right)^{\gamma_{\theta}(\alpha)/2}.
\label{L14a}
\end{eqnarray}
The vanishing of $\beta(\alpha)$ can be achieved in various ways. The function $\beta(\alpha)$ could have a non-trivial zero for some specific value of the coupling constant $\alpha$. For an Abelian gauge theory this could only occur non-perturbatively, while for a non-Abelian one, with an appropriate fermion content the second-order term in $\alpha$ could have the opposite sign to that of the first-order term, to then permit a perturbative cancellation. However, for our purposes here, one does not need to actually require that $\beta(\alpha)$ have such a zero, since one could instead work in the much studied quenched approximation in which one keeps the photon canonical  (i.e. $\beta(\alpha)$ is zero for all $\alpha$ and there is no charge renormalization). In this approximation, one obtains the asymptotic scaling form for $\tilde{S}^{-1}(p)$ if one sums all photon exchange diagrams both planar and non-planar combined \cite{Johnson1964}, doing so for any value of the coupling constant. Or one could keep quenched planar graphs alone and then \cite{Maskawa1974,Maskawa1975,Miransky1985} get scaling if $\alpha \leq\pi/3$, with the dynamical dimension $d_{\theta}(\alpha)=3+\gamma_{\theta}(\alpha)$ of  $\bar{\psi}\psi$ being given by  $d_{\theta}(\alpha)=2+(1-3\alpha/\pi)^{1/2}$.   In all of these cases the dressing of the $\bar{\psi}\psi$ vertex with quenched photons converts the point $\tilde{\Gamma}_{\rm S}(p,p,0)=1$ Green's function into the dressed $\tilde{\Gamma}_{\rm S}(p,p,0)=(-p^2/m^2)^{\gamma_{\theta}(\alpha)/2}$, and that is the only requirement that we will need for this work as it will soften the point vertices if $\gamma_{\theta}(\alpha)<0$, and lead to completely finite scalar and pseudoscalar channel fermion-antifermion scattering amplitudes if $\gamma_{\theta}(\alpha)=-1$. Since the summation of all planar plus non-planar quenched photon graphs graph leads to scaling for any value of $\alpha$, and since  $\gamma_{\theta}(\alpha)$ is a continuous function of $\alpha$, there will (presumably) be some value of $\alpha$ for which $\gamma_{\theta}(\alpha)=-1$, and in fact this is already seen in the quenched planar graph approximation if $\alpha=\pi/3$. In the following we thus explore the implications of $\gamma_{\theta}(\alpha)=-1$.

Once there is critical scaling no matter what the cause, the bare mass behaves as
\begin{eqnarray}
m_0=m\left(\frac{\Lambda^2}{m^2}\right)^{\gamma_{\theta}(\alpha)/2},
\label{L15a}
\end{eqnarray}
and thus vanishes if $\gamma_{\theta}(\alpha)<0$. In consequence, $\delta(m)=m-m_0$ is finite. With a zero $m_0$ and a non-zero $m$ this looks like dynamical symmetry breaking. However, with $Z_{\theta}^{-1/2}=(\Lambda^2/m^2)^{\gamma_{\theta}(\alpha)/2}$ effecting $Z_{\theta}^{-1/2}(\bar{\psi}\psi)_0=\bar{\psi}\psi$, $m_0(\bar{\psi}\psi)_0=m(\bar{\psi}\psi)$ is not zero. Because of this, chiral symmetry is broken in the Lagrangian, and there is no associated Goldstone boson \cite{Baker1971a}.

In the quenched planar graph approximation it was found \cite{Maskawa1974,Maskawa1975,Miransky1985} that if $\alpha > \pi/3$ the fermion propagator does not scale asymptotically and there then is a Goldstone boson. Thus the prevailing wisdom from that time on has been that the generation of dynamical Goldstone bosons is associated with strong coupling only (cf. $\alpha>\pi/3$). In this paper and the more detailed \cite{Mannheim2015,Mannheim2016} we revisit this wisdom and show that there can be a Goldstone boson even for weak coupling and critical scaling. However, since there is none in the fermion gauge boson theory itself, we will need to embed the theory in a larger chiral invariant one, namely an NJL type theory, one that would be power-counting renormalizable if  $d_{\theta}(\alpha)=2$, since then  $(\bar{\psi}\psi)^2$ would act as an operator whose dynamical dimension is reduced from six to four. (In \cite{Mannheim2015,Mannheim2016} we discuss dynamical Goldstone boson studies by other authors, some of which also involve $d_{\theta}(\alpha)=2$.)

\subsection{Non-vanishing of Physical Fermion Mass}

With the exact photon-fermion-antifermion vertex $\Gamma^{\mu}(p+q,p)$ obeying the unrenormalized $\Gamma^{\mu}(p+q,p)=\gamma^{\mu}-\int d^4k/(2\pi)^4S(p+q+k)\Gamma^{\mu}(p+q+k,p+k)S(p+k)K(p+k,p+q+k,k)$ where $S(p)=1/(\slashed{p}-m_0-\Sigma(p))$ and $K(p+k,p+q+k,k)$ is the Bethe-Salpeter kernel, from the Ward identity $q_{\mu}\Gamma^{\mu}(p+q,p)=S^{-1}(p+q)-S^{-1}(p)$, we obtain
\begin{eqnarray}
&&\Sigma(p)-\Sigma(p+q)=-\int \frac{d^4k}{(2\pi)^4} S(p+q+k)(\Sigma(p+k)
 -\Sigma(p+q+k))S(p+k)K(p+k,p+q+k,k)
\nonumber\\
&&-\int \frac{d^4k}{(2\pi)^4}\frac{[\slashed{p}+\slashed{q}+\slashed{k}+m_0+\Sigma(p+q+k)]}
{[(p+q+k)^2-(m_0+\Sigma(p+q+k))^2]}\slashed{q}\frac{[(\slashed{p}+\slashed{k}+m_0+\Sigma(p+k))]}{[(p+k)^2-(m_0+\Sigma(p+k))^2]}K(p+k,p+q+k,k).
\label{L16a}
\end{eqnarray}
Now one can find a gauge \cite{Johnson1964}, the generalized Landau gauge, in which $\Sigma(p)$ has no Dirac gamma matrix dependence. Thus in the second term in (\ref{L16a}) the only terms that will survive will be those that are linear in $m_0+\Sigma$. With the kernel being no more divergent than $1/k^2$ if the photon is canonical, then given (\ref{L14a}) and (\ref{L15a}), the $m_0+\Sigma$ terms will appear in an integral that is finite. One can thus drop the $m_0$  dependence in the second term in (\ref{L16a}). In consequence, (\ref{L16a}) becomes a self-consistent homogeneous equation for $\Sigma(p)$, and since (\ref{L16a}) involves a difference of two propagators it has better asymptotic convergence properties than the unrenormalized Schwinger-Dyson equation $\Sigma(p)=ie_0^2\int d^4k/(2\pi)^4 D_{\mu\nu}(k)\Gamma^{\mu}(p,p-k)S(p-k)\gamma^{\nu}$. As such, (\ref{L16a}) can have the asymptotic solution $\Sigma(p)=m(-p^2/m^2)^{\gamma_{\theta}(\alpha)/2}$ as given in (\ref{L14a}) above. However, since (\ref{L16a}) is homogeneous, it can also have the completely trivial solution $\Sigma(p)=0$ (i.e. both $m_0$ and $m$ identically zero). The trivial solution to (\ref{L16a}) is thus associated with the Lagrangian ${\cal{L}}_{\rm QED}^0$, while the non-trivial solution is associated with ${\cal{L}}_{\rm QED}^m={\cal{L}}_{\rm QED}^0-m\bar{\psi}\psi$.

\begin{figure}[htpb]
\begin{center}
\includegraphics[width=3.3in,height=1.3in]{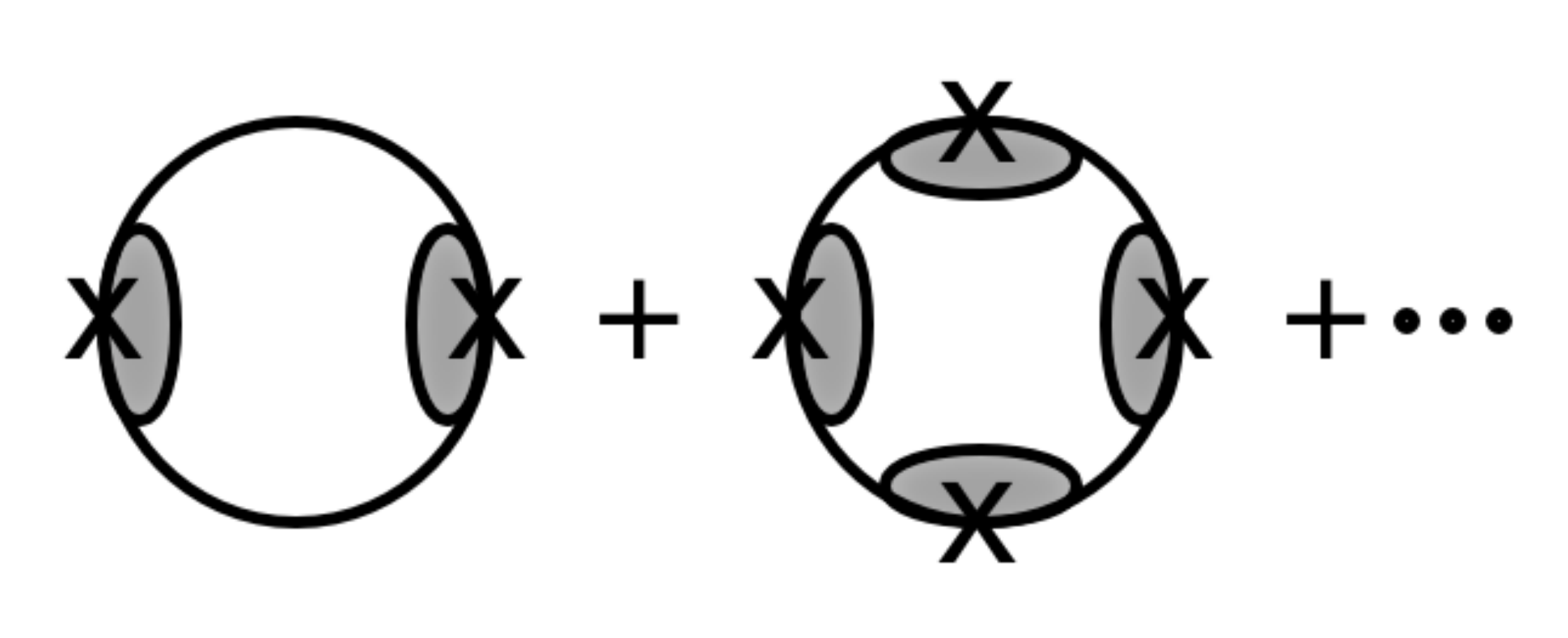}
\end{center}
\caption{Vacuum energy density  $\epsilon(m)$ via an infinite summation of massless graphs with zero-momentum dressed $m\bar{\psi}\psi$ insertions.}
\label{fig3}
\end{figure}

We thus need a criterion that would oblige us to select the non-trivial solution, and thus need to show that it has lower energy than the trivial one, i.e. just as in the NJL case,  we need to show that the $\epsilon(m)=(\langle \Omega_m|H|\Omega_m\rangle-\langle \Omega_0|H|\Omega_0\rangle)/V$ associated with $I_{\rm QED}^m$ is negative when $m\neq 0$. Now $\epsilon(m)$ is given by $\sum(1/n!)G^{(n)}_0(q_{\mu}=0,m=0)m^n$. In this sum each Green's function is associated not with a massive fermion but a massless one. Now if the massless fermion theory has critical scaling, the relations 
\begin{eqnarray}
\tilde{S}^{-1}(p,m=0)= \slashed{p},\qquad \tilde{\Gamma}_{\rm S}(p,p,0)=\left(\frac{-p^2-i\epsilon}{\mu^2}\right)^{\gamma_{\theta}(\alpha)/2}
\label{L17a}
\end{eqnarray}
will not only hold for asymptotic momenta, they will hold for all momenta in the massless theory since the massless theory has no intrinsic mass scale. (As is typical of massless theories, we renormalize Green's functions off shell at some spacelike point with $p^2=-\mu^2$, and while we can set $\mu$ equal to the physical fermion mass $M$, it is more instructive to keep $\mu$ as is and only set it equal to $M$ at the end.) To determine $\epsilon(m)$ we thus replace the $\tilde{\Gamma}_{\rm S}(p,p,0)=1$ point vertices of Fig. (\ref{fig1}) by the dressed $\tilde{\Gamma}_{\rm S}(p,p,0)$ vertices \cite{Mannheim1974,Mannheim1975,Mannheim1978} to obtain Fig. (\ref{fig3}).

\begin{figure}[htpb]
\begin{center}
\includegraphics[width=3.0in,height=1.9in]{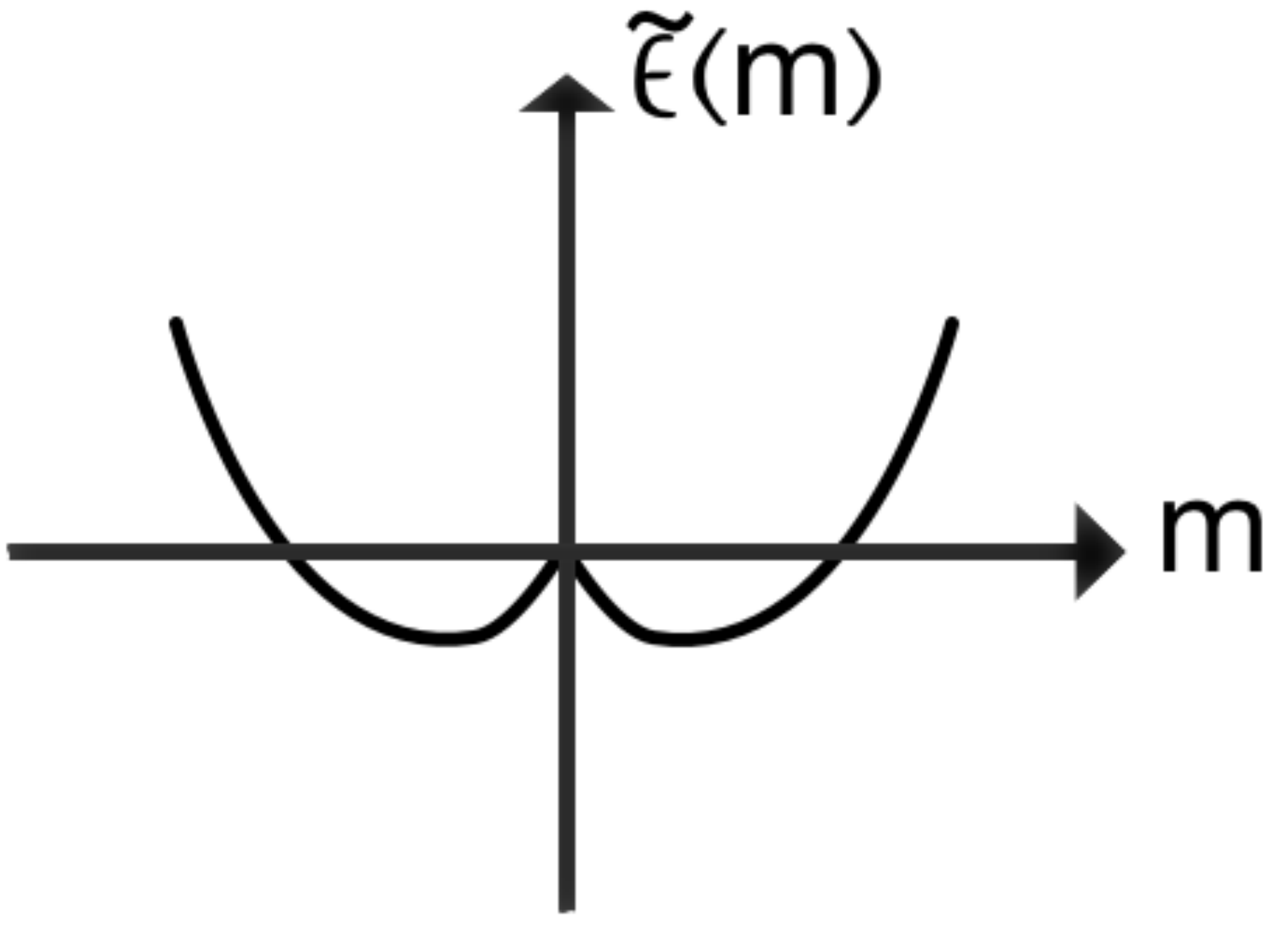}
\end{center}
\caption{Dynamically generated double-well potential for the renormalized vacuum energy density when $\gamma_{\theta}(\alpha)=-1$.}
\label{fig4}
\end{figure}

The graphs in Fig. (\ref{fig3}) can be summed analytically, and  in terms of the propagator
\begin{eqnarray}
\tilde{S}^{-1}_{\mu}(p)&=& \slashed{p}-m\left(\frac{-p^2-i\epsilon}{\mu^2}\right)^{\gamma_{\theta}(\alpha)/2}+i\epsilon,
\label{L18a}
\end{eqnarray}
 yield  \cite{Mannheim1974,Mannheim1975,Mannheim1978}
\begin{eqnarray}
\epsilon(m)=i\int \frac{d^4p}{(2\pi)^4}{\rm Tr}\left[{\rm ln}(\tilde{S}^{-1}_{\mu}(p))-{\rm ln}(\slashed{p}+i\epsilon)\right].
\label{L19a}
\end{eqnarray}
As a function of $\gamma_{\theta}(\alpha)$, $\epsilon(m)$ is found \cite{Mannheim1974,Mannheim1975,Mannheim1978} to have the shape of a single well if $-1<\gamma_{\theta}(\alpha)<0$, and to have the shape of an unbounded, upside down single well if $\gamma_{\theta}(\alpha)<-1$. However $\epsilon(m)$ is found to have the shape of a double-well potential if $\gamma_{\theta}(\alpha)=-1$, i.e. if $d_{\theta}(\alpha)$ is precisely at the value that makes $(\bar{\psi}\psi)^2$ act as a  renormalizable operator \cite{Mannheim1975}, with it taking the form
\begin{eqnarray}
\epsilon(m)=-\frac{m^2\mu^2}{16\pi^2}\left[{\rm ln}\left(\frac{\Lambda^4}{m^2\mu^2}\right)+1\right].
\label{L20a}
\end{eqnarray}
While taking $\gamma_{\theta}(\alpha)$ to be negative softens the short-distance behavior of the theory (for any negative $\gamma_{\theta}(\alpha)$ the bare mass vanishes), as $\gamma_{\theta}(\alpha)$ is taken to be more and more negative, the theory becomes more and more divergent in the infrared, and at $\gamma_{\theta}(\alpha)=-1$ the infrared divergences become so severe that the theory is forced into a new, dynamically broken, vacuum. (In the quenched planar graph approximation of \cite{Maskawa1974,Maskawa1975,Miransky1985} the condition $\gamma_{\theta}(\alpha)=-1$ is realized right at the critical $\alpha=\pi/3$ value.) As constructed, $\epsilon(m)$ is logarithmically divergent (the drop in dimension of $\bar{\psi}\psi$ from three to two converts the original quadratic divergence in the point-coupled NJL model to logarithmic). As with the mean-field sector of the NJL model, we introduce (and actually generate in ${\cal{L}}_{\rm QED-MF}$ below) a $-m^2/2g$ counterterm and set $\tilde{\epsilon}(m)=\epsilon(m)-m^2/2g$. We then identify the physical mass as the one that obeys $\tilde{\epsilon}^{\prime}(M)=\epsilon^{\prime}(M)-M/g=0$ (this relation serves to express $g$ in terms of $M$). Since  $\epsilon^{\prime}(m)=\langle \Omega_m|\bar{\psi}\psi|\Omega_m\rangle$, the physical mass and $\tilde{\epsilon}(m)$ are given by  \cite{Mannheim1974,Mannheim1975,Mannheim1978}
\begin{eqnarray}
\frac{M}{g}=\langle \Omega_M|\bar{\psi}\psi|\Omega_M\rangle=-i\int \frac{d^4p}{(2\pi)^4}{\rm Tr}[\tilde{\Gamma}_{\rm S}(p,p,0)\tilde{S}_{\mu}(p)]
=i\int \frac{d^4p}{4\pi^4}\frac{M\mu^2}{(p^2+i\epsilon)^2+M^2\mu^2}
=-\frac{M\mu^2}{4\pi^2}{\rm ln}\left(\frac{\Lambda^2}{M\mu}
\right),
\label{L21a}
\end{eqnarray}
\begin{eqnarray}
\tilde{\epsilon}(m)=\frac{m^2\mu^2}{16\pi^2}\left[{\rm ln}\left(\frac{m^2}{M^2}\right)-1\right],
\label{L22a}
\end{eqnarray}
with (\ref{L21a}) having a non-trivial solution no matter how small a negative, viz. attractive,  $g$ might be.
We recognize $\tilde{\epsilon}(m)$ as being completely finite and having the shape of the double-well potential exhibited in Fig. (\ref{fig4}). (In contrast, the point-coupled NJL $\tilde{\epsilon}(m)$ is log divergent.)
\begin{figure}[htpb]
\begin{center}
\includegraphics[width=3.4in,height=0.8in]{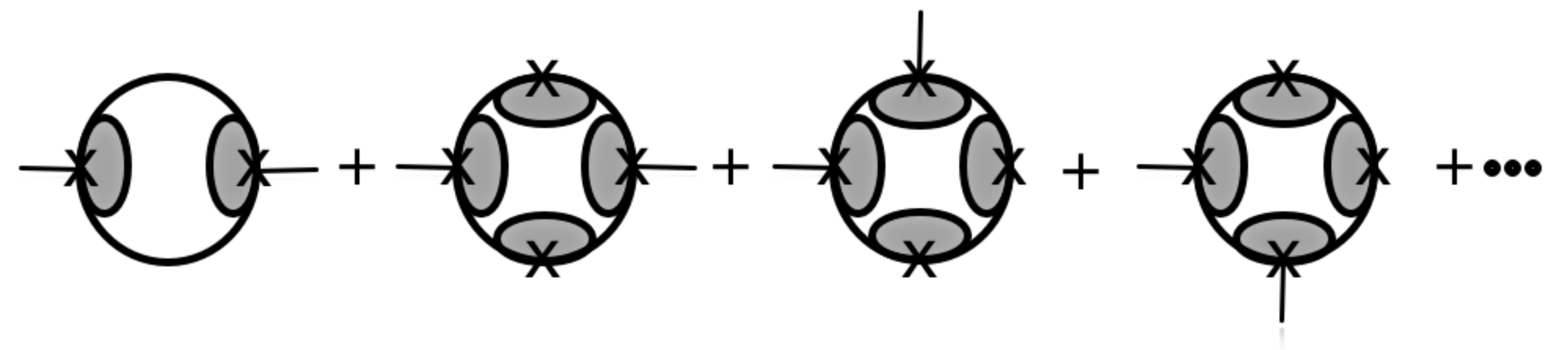}
\end{center}
\caption{$\Pi_{\rm S}(q^2,m(x))$ developed as an infinite summation of massless graphs, each with two dressed $m\bar{\psi}\psi$  insertions carrying momentum $q_{\mu}$ (shown as external lines), with all other dressed $m\bar{\psi}\psi$  insertions carrying zero momentum.}
\label{fig5}
\end{figure}

We can also parallel the NJL discussion of coherent states by taking a spacetime-dependent $m(x)$. We replace Fig. (\ref{fig2}) by its dressed version given in Fig. (\ref{fig5}). On summing the graphs in Fig. (\ref{fig5}) we obtain  \cite{Mannheim1978} 
\begin{eqnarray}
\Pi_{\rm S}(q^2,m)=-i\int \frac{d^4p}{(2\pi)^4}{\rm Tr}\bigg[\tilde{\Gamma}_{\rm S}(p+q,p,-q)
 \tilde{S}_{\mu}(p)\tilde{\Gamma}_{\rm S}(p,p+q,q)\tilde{S}_{\mu}(p+q)\bigg],
\label{L23a}
\end{eqnarray}
where $\tilde{\Gamma}_{\rm S}(p,p+q,q)=\tilde{\Gamma}_{\rm S}(p+q,p,-q)=[(-p^2/\mu^2)(-(p+q)^2/\mu^2)]^{\gamma_{\theta}(\alpha)/4}$, and thus obtain the effective Higgs Lagrangian \cite{Mannheim1978}
\begin{eqnarray}
{\cal{L}}_{\rm EFF}=-\frac{m^2(x)\mu^2}{16\pi^2}\left[{\rm ln}\left(\frac{m^2(x)}{M^2}\right)-1\right]
+\frac{3\mu}{256\pi m(x)}\partial_{\mu}m(x)\partial^{\mu}m(x),
\label{L24a}
\end{eqnarray}
together with higher-order derivative terms. (Since $m(x)$ is a c-number rather than a q-number, higher-order derivative terms do not affect renormalizability.)

As we see, at critical scaling the theory based on the Lagrangian ${{\cal L}}^0_{QED}-m\bar{\psi}\psi+m^2/2g=-(1/4)F_{\mu\nu}F^{\mu\nu} +\bar{\psi}\gamma^{\mu}(i\partial_{\mu}-eA_{\mu})\psi-m\bar{\psi}\psi+m^2/2g$ has all the trappings of dynamical symmetry breaking seen in the NJL model save only that it has no pseudoscalar or scalar bound states, with the presence of the $m\bar{\psi}\psi $ term (as expressly required by the non-vanishing of $m_0(\bar{\psi}\psi)_0$ even as $m_0$ vanishes) indicating that the chiral symmetry is broken at the level of the Lagrangian. Now in our study above of the NJL model we encountered an analogous situation, with its mean-field sector having  precisely this same non-chiral-invariant structure. Thus if we could have ${{\cal L}}^0_{QED}-m\bar{\psi}\psi+m^2/2g$ emerge in a larger theory that is chiral invariant, we could reinterpret it as the mean-field sector of that larger theory. And if we could do so, we would not expect a critical scaling  ${{\cal L}}^0_{QED}-m\bar{\psi}\psi+m^2/2g$ to possess dynamical bound states, since a mean-field sector never does. Rather, such bound states should be generated by a residual interaction, just as we now show.

\subsection{The Collective Goldstone and Higgs Modes}

To generate bound states we augment massless QED with a four-fermion interaction to give the chiral invariant Lagrangian ${\cal {L}}_{\rm QED-FF}$, and then decompose it as
\begin{eqnarray}
{\cal {L}}_{\rm QED-FF}&=&-\frac{1}{4}F_{\mu\nu}F^{\mu\nu}+\bar{\psi}\gamma^{\mu}(i\partial_{\mu}-eA_{\mu})\psi 
-\frac{g}{2}[\bar{\psi}\psi]^2-\frac{g}{2}[\bar{\psi}i\gamma^5\psi]^2
\nonumber\\
&=&-\frac{1}{4}F_{\mu\nu}F^{\mu\nu}+\bar{\psi}\gamma^{\mu}(i\partial_{\mu}-eA_{\mu})\psi 
-m\bar{\psi}\psi +\frac{m^2}{2g}
-\frac{g}{2}\left(\bar{\psi}\psi-\frac{m}{g}\right)^2-\frac{g}{2}\left(\bar{\psi}i\gamma^5\psi\right)^2
\nonumber\\
&=&{\cal{L}}_{\rm QED-MF}+{\cal{L}}_{\rm QED-RI}.
\label{L25a}
\end{eqnarray}
With $\Pi_{\rm S}(q^2,m)$ of (\ref{L23a}) being generated via Fig. (\ref{fig5}), analogously one can evaluate $\Pi_{\rm P}(q^2,m)$, to obtain 
\begin{eqnarray}
\Pi_{\rm P}(q^2,m)=-i\int \frac{d^4p}{(2\pi)^4}{\rm Tr}\bigg[\tilde{\Gamma}_{\rm S}(p+q,p,-q) i\gamma^5\tilde{S}_{\mu}(p)\tilde{\Gamma}_{\rm S}(p,p+q,q)i\gamma^5\tilde{S}_{\mu}(p+q)\bigg].
\label{L26a}
\end{eqnarray}
On translating $p_{\mu}$ to $p_{\mu}-q_{\mu}/2$, we  can thus set 
\begin{eqnarray}
\Pi_{\rm S}(q^2,m)=-4i\mu^2\int \frac{d^4p}{(2\pi)^4}\frac{N(q,p)+m^2\mu^2}{D(q,p,m)},
\qquad \Pi_{\rm P}(q^2,m)=-4i\mu^2\int \frac{d^4p}{(2\pi)^4}\frac{N(q,p)-m^2\mu^2}{D(q,p,m)},
\label{L27a}
\end{eqnarray}
where
\begin{eqnarray}
N(q,p)&=&(p^2+i\epsilon-q^2/4)(-(p-q/2)^2-i\epsilon)^{1/2}
(-(p+q/2)^2-i\epsilon)^{1/2},
\nonumber\\
D(q,p,m)&=&(((p-q/2)^2+i\epsilon)^2+m^2\mu^2)(((p+q/2)^2+i\epsilon)^2+m^2\mu^2).
\label{L28a}
\end{eqnarray}

Evaluating $\Pi_{\rm P}(q^2,m)$ at $q^2=0$ yields 
\begin{eqnarray}
\Pi_{\rm P}(q^2=0,m)&=&-4i\mu^2\int \frac{d^4p}{(2\pi)^4}\frac{(p^2)(-p^2)-m^2\mu^2}{((p^2+i\epsilon)^2+m^2\mu^2)^2}
=4i\mu^2\int \frac{d^4p}{(2\pi)^4}\frac{1}{(p^2+i\epsilon)^2+m^2\mu^2}.
\label{L29a}
\end{eqnarray}
With $m=M$ we recognize this expression as precisely being equal to $g^{-1}$. We thus find a pole at $q^2=0$ in the pseudoscalar $T_{\rm P}(q^2,M)=[g^{-1}-\Pi_{\rm P}(q^2,M)]^{-1}$, with the scattering amplitude behaving as 
\begin{eqnarray}
T_{\rm P}(q^2,M)=\frac{128\pi M}{7\mu q^2}
\label{L30a}
\end{eqnarray}
near the pole, with (unlike NJL) the residue at the pole being completely finite. Thus just as required, with dynamical symmetry breaking the residual interaction generates a massless pseudoscalar Goldstone boson that is not present in the critical scaling mean-field sector.

\begin{figure}[htpb]
\includegraphics[width=2.5in,height=2.0in]{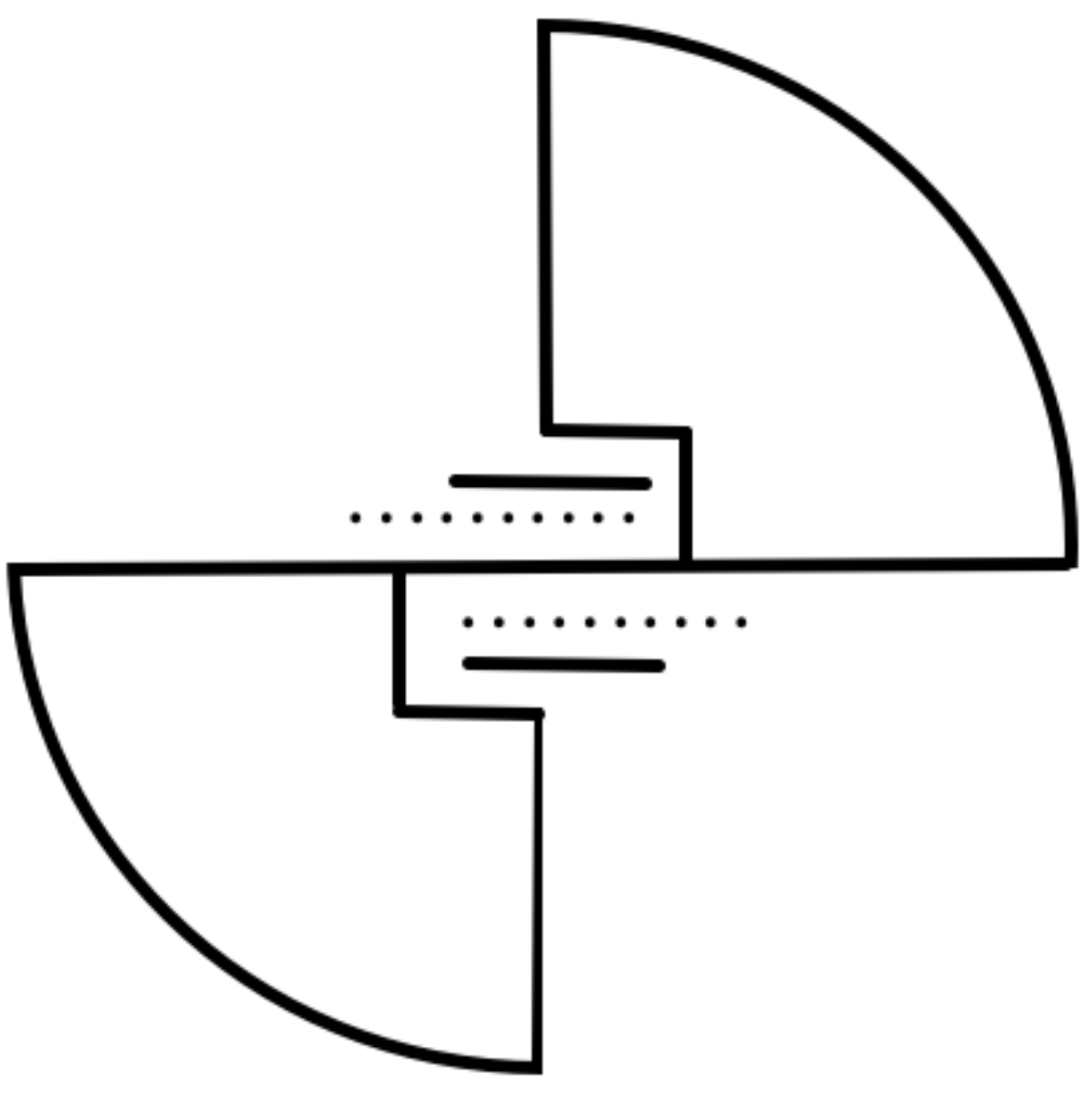}
\caption{The Wick contour in the complex $p_0$ plane. The branch cuts are shown as lines and the poles as dots.}
\label{fig6}
\end{figure}

The evaluation of $\Pi_{\rm S}(q^2,m)$ is not nearly as straightforward since we are interested in finding a pole in  $T_{\rm S}(q^2,M)$ at some non-zero $q^2$. Moreover, the analytic structure of $\Pi_{\rm S}(q^2,m)$ is much more complicated than in the point-coupled case \cite{Mannheim2015}. Specifically, if for timelike $q^2$ we set $q_{\mu}=(q_0,0,0,0)$ we find that for $(p_1^2+p_2^2+p_3^2)^{1/2}=p<q_0/2$, as well as branch points in the lower right- and upper left-hand quadrants in the complex $p_0$ plane, the function $N(q,p)$ also has branch points in the upper right- and lower left-hand quadrants. The Wick rotation for timelike $q^2$ thus has to follow the contour given in Fig. (\ref{fig6}). 
Because of the location of the branch points in $N(q,p)$, the function $\Pi_{\rm S}(q^2,m)$ is complex for all $q^2>0$, and thus poles in  $T_{\rm S}(q^2,M)=[g^{-1}-\Pi_{\rm S}(q^2,M)]^{-1}$ cannot have real $q^2$. To cancel this complex piece we will need some other complex contribution. Thus is provided by $D(q,p,m)$ as it has a branch point of its own at $q^2=2m\mu$. Solutions to $g^{-1}-\Pi_{\rm S}(q^2,M)=0$ must thus lie above the $q^2=2M\mu$ threshold. Thus once we dress the point vertices of the original point-coupled NJL model the dynamical scalar bound state pole in $T_{\rm S}(q^2,M)$ must move into the complex $q^2$ plane and become an above-threshold resonance. This provides a quite sharp contrast with an elementary Higgs boson since the mass of an elementary Higgs boson is given by the magnitude of the second derivative of the Higgs potential at its minimum, a quantity that must be real if the potential itself is.

On numerically evaluating $T_{\rm S}(q^2,M)$ \cite{Mannheim2015,Mannheim2016} a dynamical Higgs boson pole is found at $q_0=(1.48-0.02i)(M\mu)^{1/2},~q^2=(2.19-0.05i)M\mu$, i.e. just above $q^2=2M^2$ (on setting $\mu=M$). Near the pole the scattering amplitude is found to have the Breit-Wigner form
\begin{eqnarray}
T_{\rm S}(q^2,M)=\frac{46.14+1.03i}{q^2-2.22M\mu+0.05iM\mu},
\label{L31a}
\end{eqnarray}
with an expressly negative imaginary part just as required for decay, with the associated decay width being fairly narrow. Such a width could potentially serve to distinguish a dynamical Higgs boson from an elementary one. With the dynamically generated Higgs mass naturally being at the fermion mass scale rather than at some large regulator mass scale, unlike in the elementary Higgs case, there automatically is no Higgs boson hierarchy problem.

Inspection of (\ref{L31a}) shows that in the dynamical Higgs case the magnitude of the residue at the pole is completely determined, with the theory thus determining the Yukawa coupling of the Higgs boson to a fermion-antifermion pair (and likewise the Goldstone boson Yukawa coupling as per (\ref{L30a})). This stands in sharp contrast to the standard electroweak theory with its  elementary Higgs boson, where the magnitudes of Yukawa couplings have to be introduced by hand.

In the above discussion of the collective modes we have iterated the $T$ matrix as $T=g+g\Pi g+g\Pi g\Pi g+...=1/(g^{-1}-\Pi)$, by taking the Bethe-Salpeter scattering amplitude kernel to be given by $\Pi_{\rm S}(q^2,m)$ in the scalar channel and $\Pi_{\rm P}(q^2,m)$ in the pseudoscalar channel, and have found that $T_{\rm S}(q^2,M)$ and $T_{\rm P}(q^2,M)$ are both completely finite. To understand this in detail, we note in the expansion for $\Pi_{\rm S}(q^2,m)$ given in Fig. (\ref{fig5}), that when $\gamma_{\theta}(\alpha)=-1$ the only term that is ultraviolet divergent is the very first term, with this term diverging as a single logarithm. Similarly, in the expansion of $\epsilon(m)$ given in Fig. (\ref{fig3}) the only term that is ultraviolet divergent is the very first term, with this term not only also diverging as a single logarithm, but doing so with the very same coefficient as the first term in Fig. (\ref{fig5}) since the first terms in Figs. (\ref{fig3}) and (\ref{fig5}) differ only by the value of the external momentum $q_{\mu}$. With $\epsilon(m)=\sum (1/n!)G^{(n)}_0(q_{\mu}=0,m=0)m^n$, it follows that in $\epsilon^{\prime}(m)=  \sum (1/(n-1)!)G^{(n)}_0(q_{\mu}=0,m=0)m^{n-1}$ the only divergence is associated with the $n=2$ term, i.e. the $G^{(2)}_0(q_{\mu}=0,m=0)m$ term. However, in the Hartree-Fock approximation given in (\ref{L21a}) we have precisely identified $\epsilon^{\prime}(M)$ with $M/g$. In consequence, the quantity $g^{-1}-\Pi_{\rm S}(q^2,M)$ is completely finite. Since the ultraviolet behavior of $\Pi_{\rm P}(q^2,M)$ is the same as that of $\Pi_{\rm S}(q^2,M)$ (the short-distance behavior of the theory being chirally symmetric as the chiral symmetry is only broken by mass generation in the infrared), the quantity  $g^{-1}-\Pi_{\rm P}(q^2,M)$ is completely finite as well.

Thus to lowest order in $g$ in the $T$ matrix kernel, the four-fermion interaction gives cutoff-independent results, due to the softening of the $\tilde{\Gamma}_{\rm S}(p,p+q,q)$ and $\tilde{\Gamma}_{\rm P}(p,p+q,q)$ vertices from their point-coupled $1$ and $i\gamma^5$ values in the NJL model, to the  
dressed $[(-p^2/\mu^2)(-(p+q)^2/\mu^2)]^{-1/4}$ and $[(-p^2/\mu^2)(-(p+q)^2/\mu^2)]^{-1/4}i\gamma^5$ forms that they take when $\gamma_{\theta}(\alpha)=-1$. To establish the renormalizability of a four-fermion interaction with these dressed vertices, we need to dress the kernel to all orders in the four-fermion coupling constant $g$. We thus need to dress $\Pi_{\rm S}(q^2,m)$, $\Pi_{\rm P}(q^2,m)$, $\epsilon(m)$ and $\epsilon^{\prime}(m)$ with internal fermion loop graphs. When this is done, it is found \cite{Mannheim2016a} that no new divergences beyond the lowest order single logarithm are generated, with their coefficients still being equal. The all-order in $g$ renormalizability of the four-fermion theory with dressed $\tilde{\Gamma}_{\rm S}(p,p+q,q)$ and $\tilde{\Gamma}_{\rm P}(p,p+q,q)$ vertices with $\gamma_{\theta}(\alpha)=-1$ is thus established.

In the literature there have been many studies of an Abelian gluon model coupled to a four-fermion interaction. One of the first times the ${\cal {L}}_{\rm QED-FF}$ theory was explored in the literature was in \cite{Mannheim1978}, though in that paper only its mean-field aspects were studied and not the residual interaction aspects that we have presented here. One of the first studies of the residual interaction dynamics associated with ${\cal {L}}_{\rm QED-FF}$ was presented in \cite{Leung1986}, with this and other studies (as recently detailed in \cite{Mannheim2015,Mannheim2016}) finding that dynamical symmetry breaking would occur if $\alpha$ and $g$ were related according to $-g\Lambda^2=\pi^2(1+(1-3\alpha/\pi)^{1/2})^2$. And as the presence of $\Lambda$ indicates, these studies all involved a cutoff. The studies that lead to this relation differ from the study presented here in a crucial way. While these studies were based on the quenched planar graph approximation so that they did involve a fermion propagator $\tilde{S}(p)$ that scaled with an anomalous dimension just as given as per (\ref{L14a}) in our study, these studies did not implement  the Callan-Symanzik equation (\ref{L13a}) and did not use the dressed $\tilde{\Gamma}_{\rm S}(p,p,0)$ given in (\ref{L14a}), but instead took it to be given by the point-coupled $\tilde{\Gamma}_{\rm S}(p,p,0)=1$. Thus for the tadpole $\langle \Omega_M|\bar{\psi}\psi|\Omega_M\rangle$, instead of setting it equal to$-i\int d^4p/(2\pi)^4{\rm Tr}[\tilde{\Gamma}_{\rm S}(p,p,0)\tilde{S}_{\mu}(p)]$ as we did in (\ref{L21a}), in these studies it was set equal to $-i\int d^4p/(2\pi)^4{\rm Tr}[\tilde{S}_{\mu}(p)]$. Without the dressed $\tilde{\Gamma}_{\rm S}(p,p,0)$ the tadpole was not softened enough (even with $\gamma_{\theta}(\alpha)=-1$), to hence require the use of a cutoff. However, in our case the tadpole is softened sufficiently to make it renormalizable. With our study also being an all-order planar plus non-planar graph quenched photon study, we are not constrained by the planar graph $-g\Lambda^2=\pi^2(1+(1-3\alpha/\pi)^{1/2})^2$ relation, a relation that imposes a lower bound on $g$. Rather, in our case (\ref{L21a}) imposes no lower bound on $g$, with it only being required to be negative (viz. attractive). Moreover, if $\beta(\alpha)$ does have a zero, and if it is given by the fine-structure constant, $\alpha$ and $g$ could then both be weak and still lead to dynamical symmetry breaking. One is thus able to associate dynamical symmetry breaking with weak coupling, with strong coupling not being required at all.

\subsection{Distinguishing Dynamical and Elementary Higgs Bosons}

The path integral associated with the massless fermion ${{\cal L}}_{\rm QED-FF}$ is of the form
\begin{eqnarray}
Z(\bar{\eta}, \eta)=\int D[\bar{\psi},\psi]\exp\bigg{[}i\int d^4x \bigg{(}\bar{\psi}i\gamma^{\mu}\partial_{\mu}\psi
-e\bar{\psi}\gamma^{\mu}A_{\mu}\psi-\frac{g}{2}(\bar{\psi}\psi)^2+\bar{\eta}\psi+\bar{\psi}\eta\bigg{)}\bigg{]}.
\label{L32a}
\end{eqnarray}
(Not displayed are  the $A_{\mu}$ integration measure, a source term for $A_{\mu}$, or $-(g/2) (\bar{\psi}i\gamma^5\psi)^2$.) On introducing a dummy field $\sigma$ we can rewrite the path integral as 
\begin{eqnarray}
Z(\bar{\eta}, \eta)
&=&\int D[\bar{\psi},\psi,\sigma]\exp\bigg{[}i\int d^4x \bigg{(}\bar{\psi}i\gamma^{\mu} \partial_{\mu}\psi
-e\bar{\psi}\gamma^{\mu}A_{\mu}\psi-\frac{g}{2}(\bar{\psi}\psi)^2+\frac{g}{2}\left(\frac{\sigma}{g}-\bar{\psi}\psi\right)^2+\bar{\eta}\psi+\bar{\psi}\eta\bigg{)}\bigg{]}
\nonumber\\
&=&\int D[\bar{\psi},\psi,\sigma]\exp\bigg{[}i\int d^4x \bigg{(}\bar{\psi}\gamma^{\mu}i\partial_{\mu}\psi
-e\bar{\psi}\gamma^{\mu}A_{\mu}\psi-\sigma\bar{\psi}\psi +\frac{\sigma^2}{2g}
+\bar{\eta}\psi+\bar{\psi}\eta\bigg{)}\bigg{]}.
\label{L33a}
\end{eqnarray}
We recognize the path integral Lagrangian that we obtain as being of precisely the same form as the mean-field ${{\cal L}}_{\rm QED-MF}$ given above, with $\sigma(x)$ replacing $m(x)$. On introducing $\tilde{\Gamma}_{\rm S}(x)$ as the Fourier transform of $\tilde{\Gamma}_{\rm S}(p,p,0)$, on doing a path integration over the fermions we obtain an effective action in the $\sigma(x)$ sector of the form
$\int D[\sigma]\exp[i\int d^4x{{\cal L}}_{\rm EFF}]$ where ${{\cal L}}_{\rm EFF}={\rm Tr}{\rm ln}[i\slashed{\partial}_x-\int d^4x^{\prime}\sigma(x^{\prime})\tilde{\Gamma}_{\rm S}(x-x^{\prime})]$. On expanding  in derivatives of $\sigma(x)$, we recognize ${{\cal L}}_{\rm EFF}$ as being the effective Higgs Lagrangian given in (\ref{L24a}), to thus generate a kinetic energy term for $\sigma(x)$. 

However, while the path integral looks like that of an elementary scalar field, there is one key difference: there is no $J(x)\sigma(x)$ source term for $\sigma(x)$. The function $Z(\bar{\eta}, \eta)$ only depends on  the fermion sources, while if the scalar field were to be elementary the path integral would be associated with $Z(\bar{\eta}, \eta, J)$ instead. With one and the same Lagrangian, the off-shell scalar field (internal exchange and loop diagrams) contributions to Green's functions with external fermion legs as generated by either $Z(\bar{\eta}, \eta)$ or  $Z(\bar{\eta}, \eta, J)$ would be identical, with it being the all-order iteration of internal $\sigma$ exchange diagrams that would generate the dynamical Goldstone and Higgs bosons that are not present in the $\sigma$ field action itself. However, $Z(\bar{\eta}, \eta, J)$ would also allow for Green's functions with external boson legs as well. Thus elementary and dynamical Higgs bosons only differ when the Higgs field goes on shell, while not differing off shell at all. With the Higgs width being an on-shell property of the Higgs field, again we see that it is in the width of the Higgs boson that one could potentially distinguish between a dynamical Higgs boson and an elementary one.

\subsection{Conformal Gravity Treatment of the Vacuum Energy Density}

If we evaluate the zero-point vacuum energy density of a free massive fermion with equation of motion $[i\gamma^{\mu}\partial_{\mu}-m]\psi=0$ and energy-momentum tensor $T^{\mu\nu}=i\hbar \bar{\psi}\gamma^{\mu}\partial^{\nu}\psi$, from the filled negative energy sea fermion modes we obtain 
\begin{eqnarray}
\langle \Omega_m |T^{00}|\Omega_m\rangle=-\frac{\Lambda^4}{16\pi^2}{\rm ln}(\Lambda^2)+\frac{\Lambda^4}{32\pi^2}-\frac{m^2\Lambda^2}{8\pi^2}
+\frac{m^4}{16\pi^2}{\rm ln}\left(\frac{\Lambda^2}{m^2}\right)+\frac{m^4}{32\pi^2},
\label{L34a}
\end{eqnarray}
to thus generate quartic, quadratic, and logarithmic divergences. In flat spacetime one is only interested in energy-density differences, and so one can subtract off the zero-point vacuum energy density $\langle \Omega_0 |T^{00}|\Omega_0\rangle=-\Lambda^4{\rm ln}(\Lambda^2)16\pi^2 +\Lambda^4/32\pi^2$ associated with a free massless fermion, to yield the energy-density difference $\epsilon(m)$. If one then makes a further subtraction via the mean-field $\tilde{\epsilon}(m)=\epsilon(m)-m^2/2g$, one is left with a logarithmic divergence.  If, on the other hand,  one is in a critical scaling theory with $\gamma_{\theta}(\alpha)=-1$, the mass independent quartic term in $\epsilon(m)$ is not affected, while the quadratic divergence term becomes logarithmically divergent and the logarithmic divergence becomes finite. Then,  the mean-field $\tilde{\epsilon}(m)$ is completely finite.

Now if one couples to gravity, one is not actually free to simply subtract away any infinite energy density contribution since gravity couples to energy density itself and not to energy-density difference. Thus to take care of the critical scaling logarithmically divergent term we need a dynamics that would induce an appropriate counterterm. As we have seen, such a counterterm can be supplied by a four-fermion interaction, an interaction that would even be renormalizable if $\gamma_{\theta}(\alpha)=-1$. It is thus gravity that forces the four-fermion interaction upon us, with it then serving to both renormalize the mean-field energy density and generate dynamical bound states. To cancel off the quartically divergent term we would need some additional field contribution, one that would have to be due to a boson since a bosonic zero-point energy density has the opposite sign to that of a fermion. Such a contribution could be provided by a supersymmetric partner of the fermion, but when the superpartner gets a mass it generates a new quadratic divergence, to thus give an unacceptably large vacuum energy density. An alternative bosonic field would be the gravitational field itself since it is also bosonic. However, no cancellation can be achieved via standard Einstein gravity, since as a quantum theory it is not renormalizable. However, conformal gravity \cite{Mannheim2006,Mannheim2012a}, viz. gravity based on the conformal invariant action $I_{\rm W}=-\alpha_g\int d^4x (-g)^{1/2}C_{\lambda\mu\nu\kappa} C^{\lambda\mu\nu\kappa}$ where $C^{\lambda\mu\nu\kappa}$ is the Weyl conformal tensor, is renormalizable, and ghost free \cite{Bender2008a,Bender2008b}. So now the cancellation can consistently be effected. Specifically, if one defines $(-g)^{-1/2}\delta I_{\rm W}/\delta g_{\mu\nu}=-2\alpha_gW^{\mu\nu}$, the conformal gravity equations of motion take the form $-4\alpha_gW^{\mu\nu}+T^{\mu\nu}=0$, with the gravitational and matter vacuum zero-point energy densities then canceling each other, with all the vacuum contributions, including those that arise due to the change in vacuum from $|\Omega_0\rangle$ to $|\Omega_m\rangle$ taking care of each other identically. (This differs from the elementary Higgs field case because there the shift in energy density is not cancelled.) With the vacuum contributions all taking care of each other when dynamical symmetry breaking is coupled to conformal gravity, what is  observed in cosmology is only the much smaller contribution of the positive energy one-particle states (one-particle matrix elements of  $4\alpha_gW^{\mu\nu}=T^{\mu\nu}$) that lie above the vacuum, and not the contribution due the entire filled vacuum itself \cite{Mannheim2015}. In this way the cosmological constant is under control.

\vfill\eject

\end{document}